\documentclass[12pt]{article}
\usepackage[pdftex]{graphicx}
\usepackage{amsmath}
\usepackage{xcolor,soul}
\usepackage{url}
\usepackage{amssymb,esint}

\title{\bf On apparent faster-than-light behavior of moving electric fields}
\author{Germano D'Abramo\\
{\small Ministero dell'Istruzione, dell'Universit\`a e della Ricerca,}\\
{\small 00041, Albano Laziale, RM, Italy}\\
{\small E--mail: {\tt germano.dabramo@gmail.com},
  {\tt germano.dabramo@posta.istruzione.it}}}

\date{\small This is a preprint of an article published in
  {\em Eur.~Phys.~J.~Plus}\,\, 2021.\\ The final authenticated version is
  available online at:\\
https://doi.org/10.1140/epjp/s13360-021-01283-5}

\begin{document}

\maketitle

\begin{abstract}
  For every observer, however distant, the electric field of a uniformly
  moving charge is always directed away from, or points towards, the
  instantaneous present position of the charge and not away from, or towards,
  the retarded position at which the observer sees it (due to the finite speed
  of light). This fact is a well-established consequence of, among others, the
  application of the Li\'enard-Wiechert potentials, and its significance for
  fundamental physics is probably not fully appreciated. Here we show how and
  why this property has non-negligible consequences for what we take for
  granted about the relativity of simultaneity and faster-than-light
  communication. In particular, if we consider two opposite electric charges
  whose distance shrinks to zero at a constant velocity (shrinking electric
  dipole), then the cancellation of the total field seems to be instantaneous
  everywhere in space and in every inertial reference frame. A simple variant
  of the shrinking electric dipole setup appears to allow a sort of
  faster-than-light communication of information. Our results provide simple
  theoretical support to the conclusions of recent experiments on the
  propagation speed of Coulomb and magnetic fields. It would also be
  interesting to explore any possible connection between our findings and
  quantum non-locality.\\
  
  \noindent {\bf Keywords:} special relativity $\cdot$ relativity of
  simultaneity $\cdot$ electromagnetism $\cdot$ electric dipole $\cdot$
  faster-than-light propagation $\cdot$ Gauss's law  $\cdot$ non-locality
  
\end{abstract}

\section{Introduction}
\label{se1}

It is a well-known result of special relativity that nothing generated at one
point in space (mass, energy, or information) can reach another point at speed
exceeding the speed of light in vacuum $c$\footnote{Charged particles moving
inside a dielectric medium can easily have a velocity higher than that of light
in that medium. However, this is not regarded as a real superluminal motion.}.
This fact derives from the application to all bodies of the relativistic
velocity-addition formula obtained ultimately by imposing the constancy of the
speed of light in every reference frame~\cite{e05}.
Einstein directly addressed the issue of an upper limit on speeds in a paper
on special relativity published in 1907~\cite{e07}. There, he stated
that ``any assumption of the spreading of an effect with a velocity greater
than the speed of light is incompatible with the theory of relativity''. By
applying the velocity-addition formula, he showed that if one assumes the
contrary, then one ``would have to consider as possible a transfer mechanism
whose use would produce an effect which {\em precedes} the cause (accompanied
by an act of will, for example)'' [emphasis in the original]. Perhaps, this is
one of the first instances (if not the first) of causality violation associated
with the possibility of superluminal motion. In a brief calculation,
Einstein showed that if an effect propagates through a material medium faster
than light, then the interval of time $T$ needed by that effect to cover a
distance between, say, point A and point B in the material turns out to be a
negative number. This is interpreted as an instance of the ``effect which
precedes the cause''. We shall return to causality violation in
Section~\ref{se4}.

However, it is well known that wave and illumination fronts may exceed
the speed of light if they are not tied to mass or to transmitting locally
produced information: consider, for instance, an illuminated spot from a
lighthouse moving along a distant mountain wall, the propagation of shadows,
or the illumination front of any intrinsically variable source of light
(see, for instance,~\cite{ne1,ne2}). In all these cases, neither mass nor
energy (or information) originates at one end of the traveled distance and
moves to the other end.

In the present paper, however, we shall show how and why the behavior of
the electric field generated by charges moving at a constant velocity has
non-negligible consequences for what we take for granted about the relativity
of simultaneity and faster-than-light communication. In Section~\ref{se2}, we
recall the well-know and established physical fact that the electric field of
a uniformly moving charge is measured by a distant observer as always
directed away from, or pointing towards, the actual, instantaneous
position of the charge and not away from, or towards, its retarded position.
We then describe the thought experiment of a shrinking electric dipole and
show that there are situations in which the cancellation of
the dipole field is instantaneous everywhere in the surrounding space.
In Section~\ref{se3}, we propose a simple variant of the shrinking electric
dipole thought experiment, and, with the application of Gauss's law, we show
how it has non-negligible consequences for the relativity of simultaneity
and seems to allow faster-than-light communication of information.
In Section~\ref{se4}, we summarize and discuss our findings, also in the
context of recent experimental results.

\section{The shrinking electric dipole}
\label{se2}

When a point charge $q$ moves at a uniform velocity $v$, the electric field
is anywhere in space always directed away from, or points towards, the
{\em instantaneous present position} of the charge.
That means that while a distant observer sees the charge in a position that
is retarded with respect to the present position (owing to the finite speed
of light), he actually measures the field as directed away from, or pointing
towards, the actual, present position. 
\begin{figure}[t]
\includegraphics[width=10.5cm]{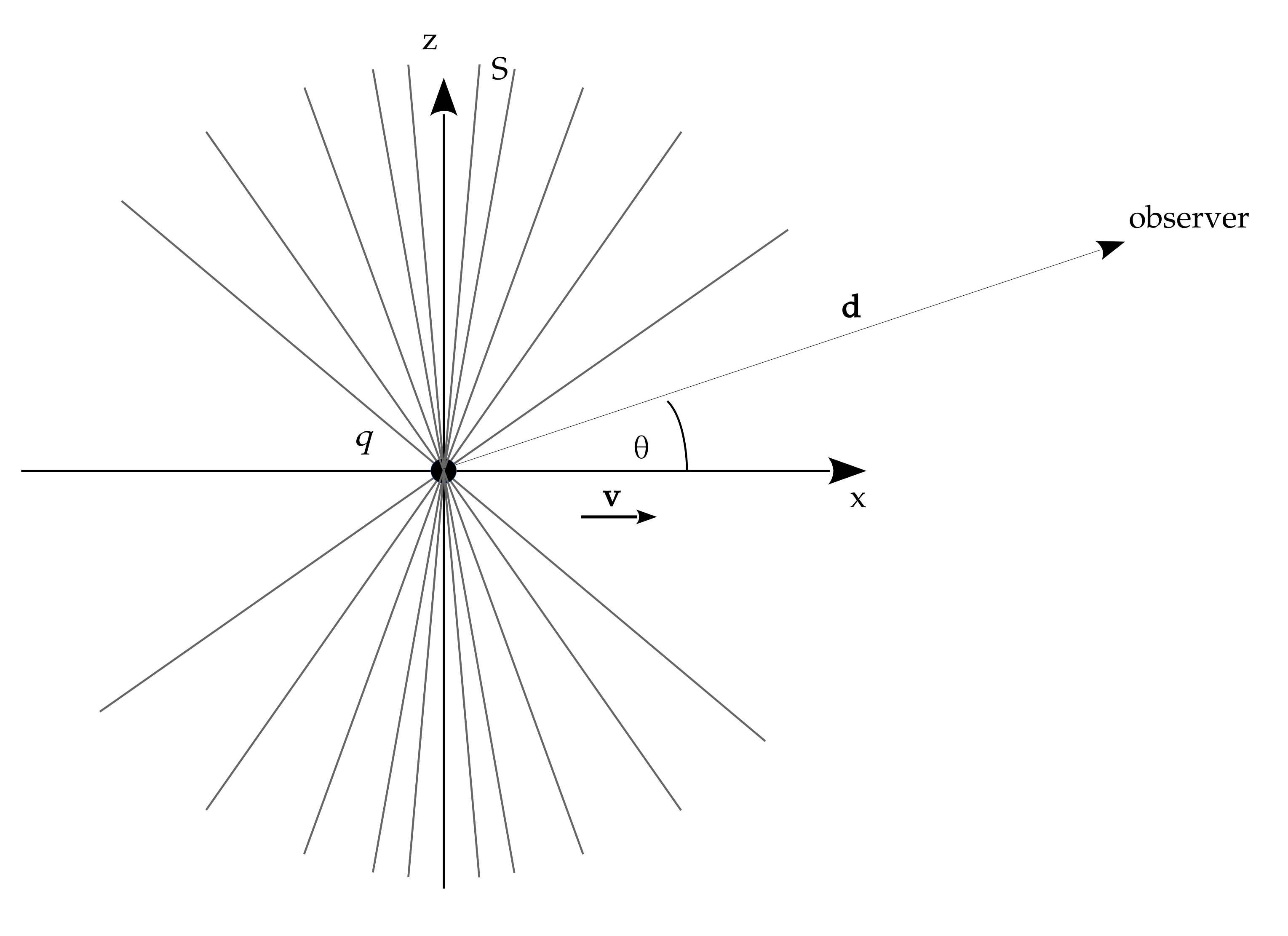}
\caption{The point charge $q$ moves at a constant velocity ${\bf v}$.
  If the magnitude of ${\bf v}$ is such that we cannot neglect the term
  $v^2/c^2$, the electric field becomes more intense at right
  angles to the motion than in the direction of the motion,
  see Eq.~(\ref{eq1}).}
\label{fig1}
\end{figure}
This peculiar feature derives directly from electromagnetism. It can be
derived from the Li\'enard-Wiechert potentials or, equivalently, from the
Lorentz transformations of fields and space-time coordinates~\cite{pur,fey,jac},
or even by appealing to the principle of relativity~\cite{dab}.
Following Purcell's or Jackson's derivations~\cite{pur,jac}, the electric
field of a point charge $q$, moving with  uniform velocity ${\bf v}$, can
be expressed in terms of the charge's instantaneous present position as

\begin{equation}
  {\bf E}=\frac{q}{4\pi\epsilon_0}\frac{1-\frac{v^2}{c^2}}{\left(1-
    \frac{v^2}{c^2}\sin^2\theta\right)^{3/2}} \frac{{\bf d}}{d^3},
\label{eq1}
\end{equation}
where $\epsilon_0$ is the vacuum permittivity, $c$ is the speed of light,
${\bf d}$ is the radial vector from the charge's present position to the
observation point, and $\theta$ is the angle between distance ${\bf d}$
and velocity ${\bf v}$ (Fig.~\ref{fig1}).

Quoting Purcell's words~\cite{pur} almost verbatim, it means that if $q$
passed the origin of the system S at precisely 12:00 p.m., S time,
an observer stationed anywhere in the system S would report that the
electric field in his vicinity was pointing, at 12:00 p.m., exactly radially
from the origin (see Fig.~\ref{fig1}).

In the following experiment, we make use of this property, and for the
sake of clarity, we refer to it as the {\em field property}. 

\begin{figure}[p]
\begin{center}
\includegraphics[width=12cm]{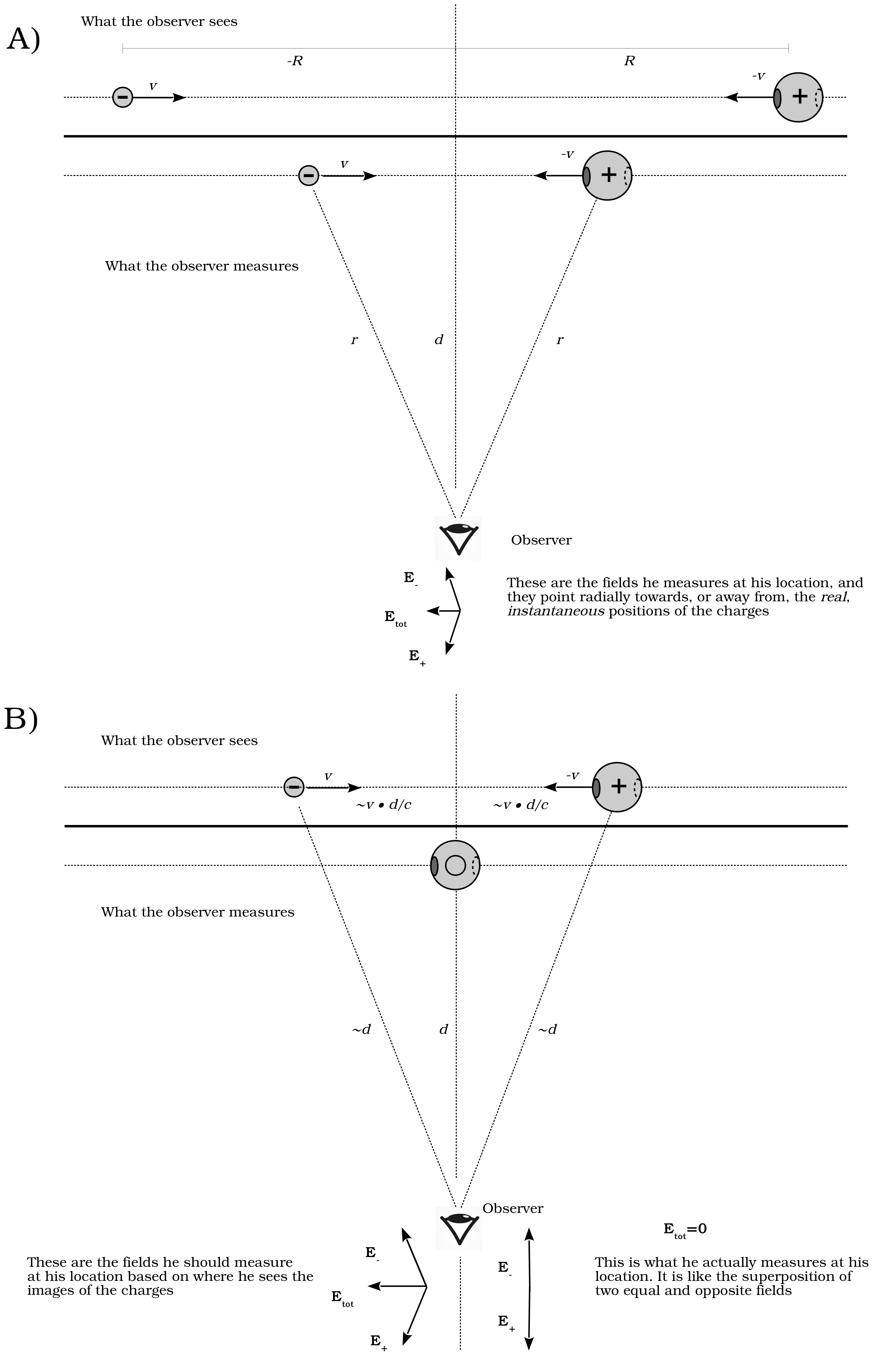}
\end{center}
\caption{Pictorial representation of the shrinking electric dipole thought
  experiment.}
\label{fig2}
\end{figure}

Consider a small metallic sphere with charge $-Q$ and a larger metallic
sphere with charge $+Q$. The larger sphere is hollow and has two small holes
along a diameter such that the smaller sphere can slide inside. These two
charge holders are separated by the distance $2R$ at initial time $t=0\, s$
and are forced to move towards each other at a constant velocity $v$ relative
to a distant, stationary observer (see Fig.~\ref{fig2}-A). If two oppositely
charged bodies were left to move freely, they would accelerate towards each
other, and the {\em field property} described in the text for uniform motion
would not hold in principle. Therefore, they are somehow forced to move at a
constant velocity.
The measure of time refers to the frame of the stationary observer. The two
charge holders generate two distinct electric fields. Furthermore, the spheres 
are oriented such that when they meet, the smaller sphere can slide completely
inside the larger one without friction and thus without suffering any
deceleration.
Let the distance between the charge holders and the observer be considerably
less than $cT$, where $T$ is the interval of time taken by the charge holders
to reach the meeting point from their starting positions. This condition
guarantees that the system is `relaxed' when the experiment starts. Any
physical information has had the time to reach the observer, and the field has
had ample time to spread all over the space where the experiment takes place.
Notice that the use of charge holders with finite spatial extension is only
functional to make the interpenetration process more clearly understandable.
According to special relativity, the geometry of moving extended bodies is
very complex, and we do not deal with this further complexity here. The reader
has to imagine all the bodies involved as being actually point-like.

The system described here represents a sort of shrinking electric dipole, and
it has already been sketched in~\cite{dab} and \cite{dab2}.

For the sake of derivation, let us assume in this Section that during the
interpenetration, the larger and the smaller sphere do not enter in
electrical contact and retain their charge.

Owing to the {\em field property} described above, the observer measures at
his location a total field that is the vector sum of the two fields directed
away from, or pointing towards, the actual, instantaneous positions of
the moving charges, no matter how distant these charges are from the
observer (Fig.~\ref{fig2}-A).

Now, consider what happens when the smaller sphere goes entirely into the
larger one and their centers overlap. Although the total charge does not go
to zero (no electrical contact), the total field in the proximity of the
sphere becomes equal to zero (superposition of equal and opposite electric
fields). 

What does happen {\em at the same instant of time} close to the distant
observer? In every phase of the process leading to the interpenetration,
the total field measured by the observer is the vector sum of the two fields
always directed away from, or pointing towards, the actual, instantaneous
positions of the moving charges (principle of superposition). Exactly at
the complete interpenetration (overlapping centers), the sum of the
two fields measured by the observer is then equal to zero. In the proximity
of the observer, it is as if we have two equal fields pointing in opposite
directions generated by two opposite and independent charges at the same
place (see Fig.~\ref{fig2}-B).

With reference to Eq.~(\ref{eq1}) and Fig.~\ref{fig2}, the total electric
field ${\bf E}_{tot}(t)$ in the proximity of the observer is equal to

\begin{multline}  
  {\bf E}_{tot}(t)= {\bf E_{-}}(t)+{\bf E_{+}}(t)=\\
  =\frac{-Q}{4\pi\epsilon_0 [(-R+vt)^2+d^2]}
  \frac{1-\frac{v^2}{c^2}}{\left\{1-\frac{v^2}{c^2}\frac{d^2}{(-R+vt)^2+d^2}
    \right\}^{3/2}}{\textrm{\bf \^{r$_{-}$}}(t)}+\\
  +\frac{+Q}{4\pi\epsilon_0 [(R-vt)^2+d^2]}
  \frac{1-\frac{v^2}{c^2}}{\left\{1-\frac{v^2}{c^2}\frac{d^2}{(R-vt)^2+d^2}
    \right\}^{3/2}}{\textrm{\bf \^{r$_{+}$}}(t)},\qquad t\in\left[0;
    \frac{R}{v}\right]
\label{eq2}
\end{multline}
where $\textrm{\bf \^{r$_{+}$}}(t)$ and $\textrm{\bf \^{r$_{-}$}}(t)$ are
the unit vectors of the distances $r_+(t)=\sqrt{(R-vt)^2+d^2}$ and
$r_-(t)=\sqrt{(-R+vt)^2+d^2}$ that separate the charge holders from
the observer,
and $\sin^2\theta(t)=\frac{d^2}{(-R+vt)^2+d^2}=\frac{d^2}{(R-vt)^2+d^2}$.

At time $T=\frac{R}{v}$, when the charge holders are at the meeting point,
the total electric field measured at the position of the observer is equal
to zero (${\bf E}_{tot}\left(\frac{R}{v}\right)={\bf 0}$). If the observer is
at any other position not equidistant from the initial positions of the
charge holders, Equation~(\ref{eq2}) still gives ${\bf E}_{tot}={\bf 0}$ at
the instant of interpenetration $T=\frac{R}{v}$ (since it is still
$r_+\left(\frac{R}{v}\right)=r_-\left(\frac{R}{v}\right)$ and
$\sin^2\theta_+\left(\frac{R}{v}\right)=
\sin^2\theta_-\left(\frac{R}{v}\right)$).

However, at the same instant of time, the charge holders are {\em seen} at
positions that are still $\frac{\frac{v}{c}d}{\sqrt{1-\frac{v^2}{c^2}}}\approx
\frac{v}{c}d$ (for $v\ll c$) away from the meeting point. This is because,
at time $T=\frac{R}{v}$, the observer is still receiving the light (the image
of the charge holders) emitted nearly $\frac{d}{c}$ time before (the exact
value is $\frac{d}{c}/\sqrt{1-\frac{v^2}{c^2}}$).

That means that the observer will {\em instantaneously} measure the
cancellation of the total field, and thus the apparent zeroing of the total
charge, no matter where or how distant this observer may be from the charges.
And, it is so despite him still seeing (i.e.,~receiving the image of) the
charges as separated in their retarded positions.

\section{Moving electric fields and faster-than-light signaling}
\label{se3}

It is possible to make a simple addition to the above shrinking electric
dipole setup with which we could transmit information faster than light.

\begin{figure}[p]
\begin{center}
\includegraphics[width=13.5cm]{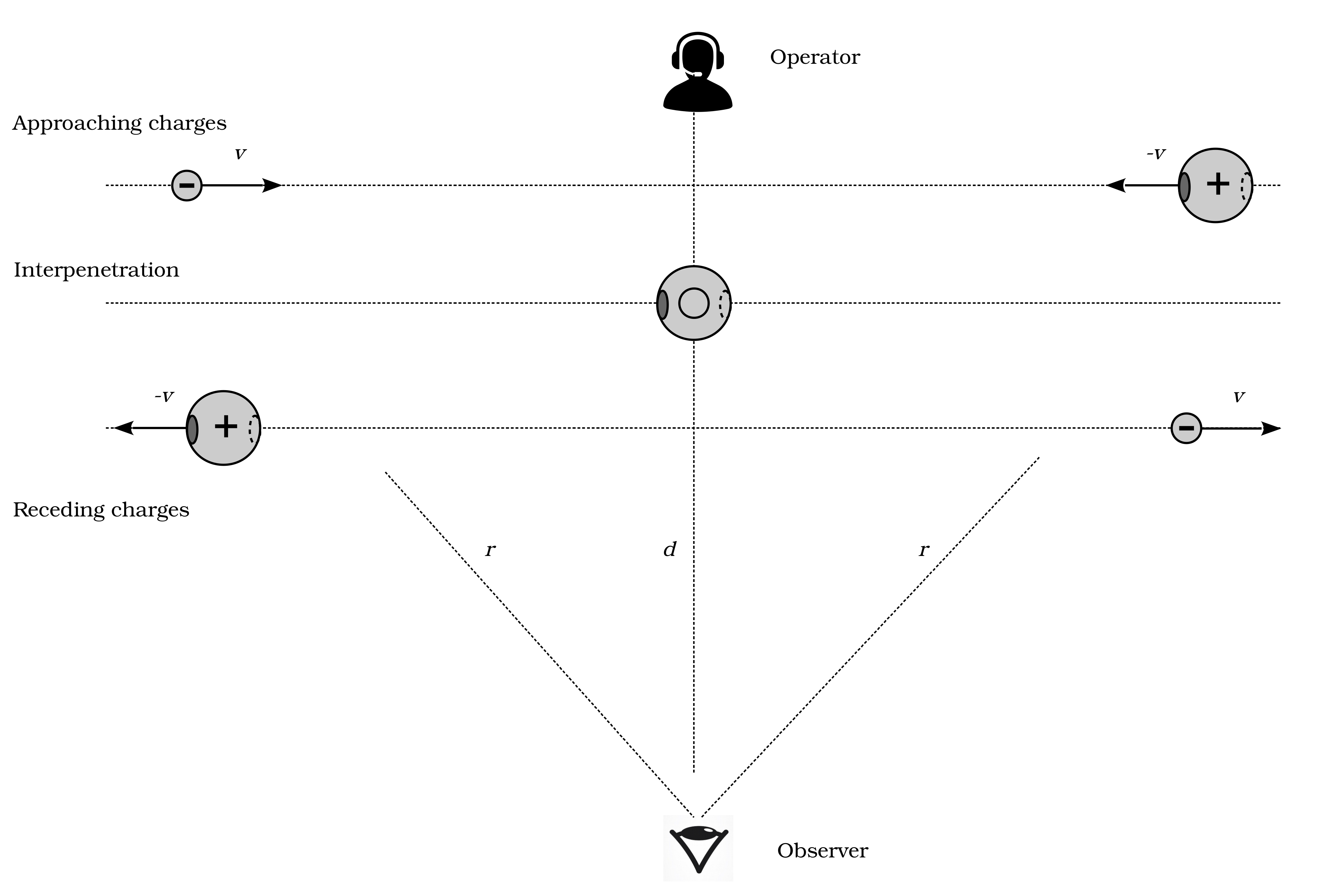}
\end{center}
\caption{Pictorial representation of the shrinking dipole variant proposed in
  the text as a tool for the faster-than-light transmission of information.
  Whether or not the operator decides to neutralize the overall charge at the
  interpenetration point (through electrical contact), owing to the
  superposition of equal and opposite electric fields, the
  total charge and field are always perceived to be equal to zero at the exact
  instant of complete interpenetration (when the two spheres' centers overlap).}
\label{fig3}
\end{figure}

Imagine the same setup as in Fig.~\ref{fig2}, this time with a human operator
close to the meeting point. In what follows, we shall also consider what
happens after the instant of interpenetration: since there is ideally no
friction, and since the smaller and the larger sphere are still forced to
move at a constant velocity, after the interpenetration, they recede from
one another at a constant velocity. Moreover, the operator, at his own will,
can allow or not allow the electrical contact between the two metallic charge
holders during the interpenetration (see Fig.~\ref{fig3}). Notice that the
operator makes the electrical contact (and thus the charge neutralization)
happen only when the smaller sphere is entirely inside the larger one (with
overlapping centers), and thus the contact happens when, from the outside,
the perceived total charge is already equal to zero (superposition of equal
and opposite electric fields).
If the operator decides to allow the electrical contact, the total charge
cancels out, and the total field is equal to zero from that moment on (at
least in the proximity of the operator). On the other hand, if the operator
decides not to allow the electrical contact, the field is equal to zero only
during the instant of time in which the smaller sphere is entirely inside the
larger one (and their centers overlap), and goes back to being different from
zero (and inverted) when the smaller and the larger sphere recede still at a
(forced) constant velocity.

Now, consider what the observer at distance $d$ from the meeting point
measures. It should be clear that the distant observer is informed
{\em instantaneously} on the operator's decision: if the observer measures a
definitive cancellation of the field, the operator has decided to allow the
electrical contact. If instead, the cancellation of the field is not definitive
(it goes to zero and then increases again in the opposite direction), the
operator has decided not to allow the electrical contact. And, according to
the analysis of the shrinking electric dipole made in Section~\ref{se2}, the
observer becomes aware of both these scenarios at the same instant of time in
which they happen close to the operator and thus without any information lag.

There is an objection to this result, though, that can be readily
advanced. It can be objected that even in the case in which the operator
chooses to allow the electrical contact, the distant observer will
see at first an instantaneous cancellation of the field (owing to the
{\em field property}), but soon he will measure the rising of an opposite
field as if the two charge holders, still moving after the interpenetration,
had not been set in contact and retained their original charge. The
information of the electrical contact (like, for instance, the image of the
operator setting the charge holders in contact) needs a time nearly equal to
$d/c$ to reach the distant observer. Only after that time, the observer would
measure the definitive cancellation of the field.

This objection follows the common interpretation of the results coming
from the application of the Li\'enard-Wiechert potentials to derive the
{\em field property} described in Section~\ref{se2}~\cite{pur, fey, car}.
Since the potentials depend only on what the charges are doing at the retarded
time, when the charge holders start to recede from the meeting point, the
potentials (and thus the fields) at the distant location of the observer
will be the same whether the charge holders recede with the same charge or
whether they recede charge-less due to the electrical contact after the
interpenetration. And, this situation will last until the information
front of the electrical contact (traveling at the speed of light) reaches
the observer.

However, we shall show that this objection is incompatible with Gauss's law.
Gauss's law holds with every closed surface, at any instant of time, and
also when charges are moving (see, for instance, Sections 5.3 and 5.4 of
reference~\cite{pur}).

\begin{figure}[p]
\begin{center}
\includegraphics[width=13.5cm]{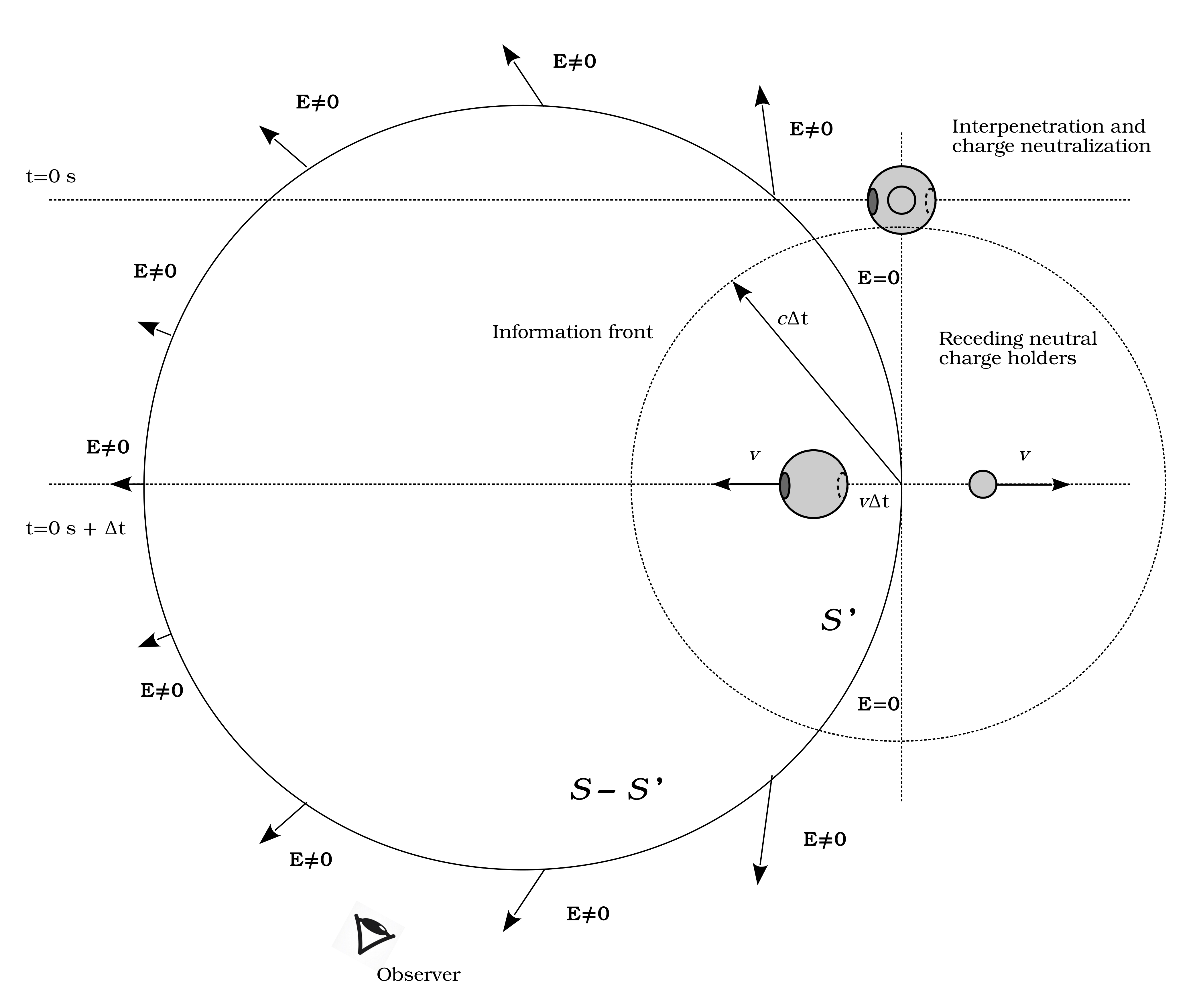}
\end{center}
\caption{At the interpenetration point (instant $t=0\,$s), the charge
  holders become neutral. After a time $\Delta t$, the information front
  of the neutralization has reached a distance $c\Delta t$. To apply Gauss's
  law, we choose a closed spherical surface $S$ such that $S$ is tangent to
  the interpenetration point, and its center lies on the trajectory of the
  charge holders (with a diameter at least as big as the distance $d$ of
  the distant observer). According to the received interpretation, although
  the charge holders are now electrically neutral, the electric field
  $\textrm{\bf E}$ should still be different from zero beyond the information
  front, and corresponding to the dipole field generated by the charge
  holders as if they retained their original charge.}
\label{fig4}
\end{figure}

Suppose that the operator decides to allow the electrical contact (overall
charge neutralization). Let us consider a closed spherical surface $S$ tangent
to the interpenetration point, with the center lying on the trajectory of the
charge holders and with a diameter at least as big as the distance $d$ of the
distant observer (see Fig.~\ref{fig4}).
If a time $\Delta t$ has passed after the interpenetration (and the overall
charge neutralization), the information that the field is now definitely equal
to zero has allegedly reached the distance $c\Delta t$ from the
interpenetration point.
Suppose that $c\Delta t\leq d$. That means that if we consider the intersection
of the spherical surface $S$ and the surface of the information front
with radius $c\Delta t$ and centered at the interpenetration point, the portion
$S'$ of $S$ that is inside the sphere of radius $c\Delta t$ has no electric
field on it ($\textrm{{\bf E}}=\textrm{{\bf 0}}$). The remaining part of $S$,
which we call $S-S'$, is instead still crossed by the field that allegedly
should be there since the information of the definitive cancellation is not
yet arrived.

The flux of the electric field $\textrm{{\bf E}}$ across the closed surface
$S$ is 

\begin{equation}
 \Phi_{\textrm{\bf E}}=\oiint_{S} \textrm{\bf E}\cdot \mathrm{d}\textrm{\bf A},
  \label{eq3}
\end{equation}
where $\mathrm{d}\textrm{\bf A}$ is a vector representing an infinitesimal
element of area of the surface $S$, and $\cdot$ represents the dot product of
two vectors. After the time $\Delta t$, the electric field $\textrm{\bf E}$
is allegedly non-zero only across the portion $S-S'$ of the total surface $S$
($S-S'$ is an open surface).

Moreover, and most importantly, the dipole field $\textrm{\bf E}$ is always
only outgoing or incoming across $S$, depending upon the sign of the former
charge on the charge holder inside it. Since, by construction, $S$ encloses
only the portion of space where there is allegedly only the positive or the
negative charge of the dipole, the sign of the dot product
$\textrm{\bf E}\cdot \mathrm{d}\textrm{\bf A}$ is always positive or negative,
respectively, for every $\mathrm{d}\textrm{\bf A}$ on the surface $S$.

Now, according to Gauss's law, the flux in Eq.~(\ref{eq3}) must be equal to
zero since there is no charge inside $S$,

\begin{equation}
  \oiint_{S} \textrm{\bf E}\cdot \mathrm{d}\textrm{\bf A}=
  \iint_{S-S'} \textrm{\bf E}\cdot \mathrm{d}\textrm{\bf A}
  =\frac{Q_{\textrm{inside $S$}}}{\epsilon_0}=0.
  \label{eq4}
\end{equation}

Notice that, at the considered instant of time, the part of the first integral
in Eq.~(\ref{eq4}) calculated on $S'$ is equal to zero.

Because the sign of the dot product $\textrm{\bf E}\cdot \mathrm{d}\textrm{\bf
A}$ is always the same for every $\mathrm{d}\textrm{\bf A}$ on the surface $S$,
the only solution to Eq.~(\ref{eq4}) is that the electric field
$\textrm{\bf E}$ is identically zero at every point on $S$.

Since both $S$ and $\Delta t$ have been chosen arbitrarily, in the case of the
operator allowing the electrical contact, the dipole field $\textrm{\bf E}$
becomes equal to zero at the moment of interpenetration and remains equal to
zero from then on for every observer in space, however distant. No information
lag is thus possible.

This result has a non-trivial, but not entirely unexpected either, impact on
the relativity of simultaneity.

According to special relativity, the simultaneity of two events occurring at
two distinct places depends upon the observer's reference frame. If one event
occurs at point $x_1$ at time $t_0$ and the other event at $x_2$ and $t_0$ (the
same time and the same reference frame), we find that the two corresponding
times $t'_1$ and $t'_2$ in the reference frame of a moving observer differ by
an amount

\begin{equation}
t'_2-t'_1=\frac{v(x_1-x_2)/c^2}{\sqrt{1-v^2/c^2}},
  \label{eq5}
\end{equation}
where $v$ is the velocity of the observer relative to the reference frame
where points $x_1$ and $x_2$ are at rest. That derives from a straightforward
application of the Lorentz transformation of the time coordinate.

The conclusion reached with Gauss's law proves, however, that the cancellation
of the dipole field is an event simultaneous at every point in space and for
every observer, regardless of the inertial reference frame in which the
observer is at rest.

Consider two points, A and B, separated in space and at rest relative to
the observer and the center of mass of the dipole (Fig.~\ref{fig2}). At the
instant in which the dipole `shrinks to zero', and there is the electrical
contact, the definitive cancellation of the field is instantaneous at
every point in space, and thus the cancellation of the field in A and B is
simultaneous for the observer.

If now the observer moves at a constant velocity relative to A and B and the
center of mass of the dipole, the application of Gauss's law is still possible
in the reference frame of the observer.
In that frame, the charge holders move no longer head-on and with equal and
opposite velocities, but, as mentioned before, Gauss's law also holds when
charges move arbitrarily. As happens in Fig.~\ref{fig4}, Gauss's law
guarantees that, for the observer, the dipole field becomes definitively
equal to zero {\em just after} the interpenetration and the electrical contact
between the charge holders, and it happens instantaneously {\em at every
point in space}. Thus, even for the moving observer, the cancellation of the
field is simultaneous at points A and B.

Notice that Gauss's law can be equivalently applied to show that if two equal
and opposite charges are generated (e.g.,~through triboelectric charging) and
separated, then the electric field of each charge should come into existence
instantaneously at every point in space. Even in this case, no information
lag is consistent with Gauss's law.

\begin{figure}[p]
\begin{center}
\includegraphics[width=13.5cm]{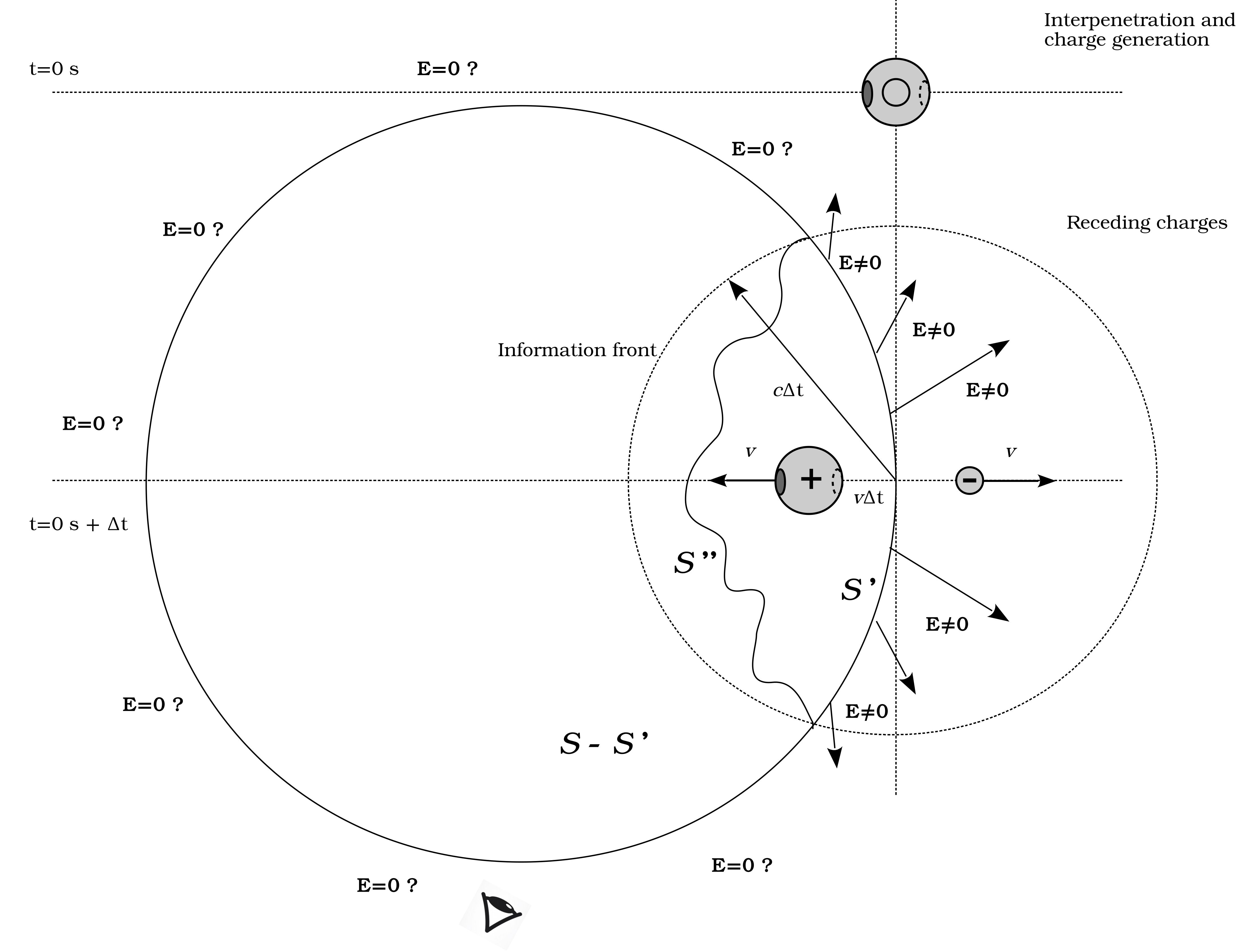}
\end{center}
\caption{At the interpenetration point (instant $t=0\,$s), the charge holders
  acquire a charge $Q$ through triboelectric charging. After a time $\Delta t$,
  the information that the electric field is now different from zero has
  reached a distance $c\Delta t$. To apply Gauss's law, we choose a closed
  spherical surface $S$ such that $S$ is tangent to the interpenetration point,
  and its center lies on the trajectory of the charge holders (with a diameter
  at least as big as the distance $d$ of the distant observer). Furthermore, an
  open surface $S''$ is considered that is entirely inside the information
  front and has as a boundary the (closed) intersection line between surface
  $S$ and the information sphere of radius $c\Delta t$. Surface $S''$, together
  with $S'$, makes a closed surface that includes the charged sphere.}
\label{fig5}
\end{figure}

Consider the situation depicted in Fig~\ref{fig5}. A smaller and a larger
sphere, both neutral, meet at the interpenetration point. They move at a
constant velocity. After the interpenetration, they recede from each other
at the same (controlled) constant velocity and with an equal and opposite
electric charge (again, the larger sphere with $+Q$ and the smaller one with
$-Q$) acquired through triboelectric charging.

Does the dipole field they produce come into existence instantaneously at
every point in space (no lag)? The answer is yes, and it comes from the
application of Gauss's law. 

Consider a closed spherical surface $S$ tangent to the interpenetration point,
with the center lying on the trajectory of the positively charged sphere and
with a diameter at least as big as the distance $d$ of the distant observer.
If a time $\Delta t$ has passed after the interpenetration (and the charging
of the sphere), the information that the field is now different from zero has
allegedly reached a distance $c\Delta t$ from the interpenetration point.
Suppose that $c\Delta t\leq d$. That means that if we consider the
intersection of the spherical surface $S$ and the surface of the information
front with radius $c\Delta t$ and centered at the interpenetration point,
{\em only} the portion $S'$ of surface $S$ inside the sphere of radius $c\Delta
t$ has an electric field different from zero on it ($\textrm{{\bf E}}\neq
\textrm{{\bf 0}}$). The remaining part of $S$, which we call $S-S'$, is not
yet crossed by the field. The field cannot be there since the information of
its generation is not arrived (see Fig.~\ref{fig5}).

Now, according to Gauss's law, the flux of the dipole field $\textrm{{\bf E}}$
must be equal to the charge inside $S$ divided by the vacuum permittivity
$\epsilon_0$

\begin{equation}
  \oiint_{S} \textrm{\bf E}\cdot \mathrm{d}\textrm{\bf A}=
  \frac{Q_{\textrm{inside $S$}}}{\epsilon_0}.
  \label{eq6}
\end{equation}

Notice that, at the considered instant of time, the integral in Eq.~(\ref{eq6})
is non-zero only on the portion $S'$ of the whole surface $S$,

\begin{equation}
  \oiint_{S} \textrm{\bf E}\cdot \mathrm{d}\textrm{\bf A}=
  \iint_{S'} \textrm{\bf E}\cdot \mathrm{d}\textrm{\bf A}
  =\frac{Q_{\textrm{inside $S$}}}{\epsilon_0}.
  \label{eq7}
\end{equation}

Consider a second open surface $S''$ that is entirely inside the information
front and has as a boundary the (closed) intersection line between surface
$S$ and the information sphere of radius $c\Delta t$. Then, $S''$ together
with $S'$ make a closed surface that includes the charged sphere
(see Fig.~\ref{fig5}). We have then

\begin{equation}
 \oiint_{S'+S''} \textrm{\bf E}\cdot \mathrm{d}\textrm{\bf A}=
  \frac{Q_{\textrm{inside $S$}}}{\epsilon_0}.
  \label{eq8}
\end{equation}

Focusing on the first members of Eqs.~(\ref{eq6}) and~(\ref{eq8}), we can
write

\begin{multline}
  \oiint _S\textrm{\bf E}\cdot \mathrm{d}\textrm{\bf A}=
  \oiint_{S'+(S-S')} \textrm{\bf E}\cdot \mathrm{d}\textrm{\bf A}=
  \iint_{S'} \textrm{\bf E}\cdot \mathrm{d}\textrm{\bf A}+
  \iint_{(S-S')} \textrm{\bf E}\cdot \mathrm{d}\textrm{\bf A}=\\
=\oiint_{S'+S''} \textrm{\bf E}\cdot \mathrm{d}\textrm{\bf A}=
  \iint_{S'} \textrm{\bf E}\cdot \mathrm{d}\textrm{\bf A}+
  \iint_{S''} \textrm{\bf E}\cdot \mathrm{d}\textrm{\bf A},
\label{eq9}
\end{multline}
and thus

\begin{equation}
  \iint_{(S-S')} \textrm{\bf E}\cdot \mathrm{d}\textrm{\bf A}=
  \iint_{S''} \textrm{\bf E}\cdot \mathrm{d}\textrm{\bf A}.
    \label{eq10}
\end{equation}

Since the flux of the dipole field through the surface $S''$ cannot be equal
to zero ($S''$ is inside the information front, and the field is non-zero
there), the same must be for the flux of the electric field through the
surface $S-S'$. Since the choice of $S$ is arbitrary, the only possibility for
$\iint_{(S-S')} \textrm{\bf E}\cdot\mathrm{d}\textrm{\bf A}$ to be different
from zero is that the electric field $\textrm{\bf E}$ is already different
from zero (and equal to that generated by the dipole) on the whole surface
$S$ {\em before} the arrival of the information front.

With reference to Fig.~\ref{fig5} and to Eqs.~(\ref{eq1}) and~(\ref{eq2}) (and
as far as Eq.~(\ref{eq1}) holds), the electric field ${\bf E}_{tot}(t)$ {\em
instantaneously} generated and measured at point ${\bf d}$ in space is given
by

\begin{equation}
  \begin{array}{llll}
    {\bf E}_{tot}(0) & = & {\bf 0} & \textrm{since $Q=0$ at $t=0$,}\\
      {\bf E}_{tot}(t) & = & \frac{-Q}{4\pi\epsilon_0 |{\bf d}-
       t\cdot {\bf v_{-}}|^2}\frac{1-\frac{v_-^2}{c^2}}{\left(1-\frac{v_-^2}{c^2}
        \sin^2\theta_-(t)\right)^{3/2}}\frac{{\bf d}-
      t\cdot {\bf v_{-}}}{|{\bf d}-t\cdot {\bf v_{-}}|} +& \\
      & & +\frac{+Q}{4\pi\epsilon_0 |{\bf d}-
        t\cdot {\bf v_{+}}|^2}\frac{1-\frac{v_+^2}{c^2}}{\left(1-\frac{v_+^2}{c^2}
        \sin^2\theta_+(t)\right)^{3/2}}\frac{{\bf d}-
     t\cdot {\bf v_{+}}}{|{\bf d}-t\cdot {\bf v_{+}}|}&  \textrm{ for $t > 0$,}
  \end{array}  
\label{eq11}
\end{equation}
where ${\bf d}$ is the distance from the center of the splitting dipole to the
point where the field is measured, ${\bf v_-}$ and ${\bf v_+}$ are respectively
the velocities of the negative and positive charge measured in the
reference frame of the observer, and
$\sin^2\theta_\pm(t)=\left(\frac{|{\bf v_\pm}\times
  ({\bf d}-t\cdot{\bf v_\pm})|}{|{\bf v_\pm}||{\bf d}-
  t\cdot{\bf v_\pm}|}\right)^2$. 

When the center of mass of the splitting dipole is stationary with the
reference frame of the observer, we have that ${\bf v_{+}}=-{\bf v_{-}}$. If
instead, the observer moves with velocity ${\bf u}$ relative to the center of
mass of the  splitting dipole, then ${\bf v_{+}}\neq-{\bf v_{-}}$ since velocity
${\bf u}$ must be subtracted from the equal and opposite velocities of the
charges measured in the center of mass of the dipole.

The above analysis suggests that field lines exist in space or do not exist.
They cannot be existent in a finite portion of space and non-existent in the
remaining (infinite) portion of space. For instance, for emission and
absorption of charged particles, Dirac explicitly writes~\cite{dir}:
``Whenever an electron is emitted, the Coulomb field around it is
simultaneously emitted, forming a kind of dressing for the electron. Similarly,
when an electron is absorbed, the Coulomb field around it is simultaneously
absorbed''.

\section{Discussion and conclusions}
\label{se4}

In the previous Sections, we have seen that the electric field of a
uniformly moving charge is measured by a distant observer as always directed
away from, or pointing towards, the instantaneous present position of the
charge and not towards or away from its retarded position. We have shown
how this fact has important consequences for the relativity of simultaneity
and the theoretical feasibility of a faster-than-light communication of
information.

If we consider an electric dipole with inter-charge distance shrinking to
zero at uniform velocity, then the cancellation of the total field seems to be
instantaneous everywhere in space and in every reference frame.
Moreover, a simple variant of the shrinking electric dipole thought experiment
appears to allow faster-than-light (actually, instantaneous) communication of
information.

Notice that faster-than-light communication of information does not violate
causality {\em per se}. It does so only if simultaneity is relative. The
relativity of simultaneity implies that time passes at different rates in
reference frames in relative motion. Then, instantaneous communication of
information may lead to the paradox of receiving a message from an observer
in motion relative to you before you ask him to send the message to you (this
situation is particularly evident when represented in a Minkowski diagram).
However, a consequence of what we have found in Section~\ref{se3} is that
simultaneity appears to be not relative, at least with the cancellation or
the generation of the field of two interpenetrating or separating equal and
opposite charges. In general, if simultaneity is not relative, then time
flows at the same pace in every reference frame. In that case,
faster-than-light or even instantaneous signaling cannot let the effect
precede the cause in any reference frame.

The theory behind the present results is well-known and well-established, and
the derivation is simple and direct. All this works in favor of our findings
in a twofold way. First, unless the physics theory behind it is fundamentally
flawed, our results should be corresponding to physical facts. Second,
owing to the underlying simplicity of our derivation, its conclusions can be
unambiguously tested in the laboratory, at least in principle.

The amplitude of a standard e.m.~wave depends on the distance from the source
as $1/r$, and thus the intensity of the wave scales as the square of the
amplitude ($1/r^2$). However, in all the examples previously made, we deal
with a quasi-static dipole field. The amplitude of the quasi-static dipole
fields depends on the distance as $1/r^3$. 
The considerable decrease in the amplitude with distance might prevent
this field from being easily detected if not purposely searched for.

Concerning already performed experimental tests, our findings, and above
all what has been derived with the application of Gauss' law in
Section~\ref{se3}, seem to provide simple theoretical support to the
conclusions of the experiments conducted by the Frascati Group on the
propagation speed of Coulomb fields~\cite{fra,fra2,ru} and by
Kholmetskii~{\em et al.}~on the propagation speed of magnetic
fields~\cite{kho1, kho2}.
Moreover, as already suggested in~\cite{kho1, kho2} for those results, it
would be interesting to explore any possible connection between our findings
and quantum non-locality. In particular, if the present results are sound and
experimentally confirmed, it would be of some interest to know whether they
may have a role in the apparent action-at-distance behavior of many quantum
experiment results (i.e.,~instantaneous correlations between the properties
of remote systems).

\section*{Acknowledgments}
The author is indebted to Dr.~Assunta Tataranni and Dr.~Gianpietro Summa for
key improvements to the manuscript.

\end{document}